\documentclass[12pt,preprint]{aastex}

\usepackage{graphicx}

\shorttitle{Prominences observations with SDO/AIA}
\shortauthors{Parenti et al.}

\begin{document}


\title{On the nature of prominence emission observed by SDO/AIA.}


\author{S. Parenti }
\affil{Royal Observatory of Belgium, 1180 Bruxelles, Be}
\email{s.parenti@oma.be}

\author{B. Schmieder}
\affil{Observatoire de Paris, LESIA, 92195 Meudon, France}

\author{P. Heinzel}
\affil{Astronomical Institute, Academy of Sciences of the Czech Republic, 25165 Ond\v rejov, Czech Republic}

\and
\author{L. Golub}
\affil{Harvard-Smithsonian Center for Astrophysics, Cambridge, MA 02138, USA}




\begin{abstract}

The Prominence-Corona Transition Region (PCTR) plays a key role in the thermal and pressure equilibrium of solar prominences. Our knowledge of this interface is limited and several major issues remain open, including the thermal structure and, in particular, the maximum temperature of the detectable plasma. The high signal-to-noise ratio of images obtained by the Atmospheric Imaging Assembly (AIA) on NASA's Solar Dynamics Observatory clearly show that prominences are often seen in emission in the 171 and 131 bands. We investigate the temperature sensitivity of these AIA bands for prominence observation, in order to infer the temperature content in an effort to explain the emission. 

Using the CHIANTI atomic database and previously determined prominence differential emission measure distributions, we build synthetic spectra to establish the main emission-line contributors in the AIA bands. We find that the \ion{Fe}{9}  line always dominates the 171 band, even in the absence of plasma at $> 10^6 ~\mathrm{K}$ temperatures, while the 131 band is dominated by \ion{Fe}{8}. We conclude that the PCTR has sufficient plasma emitting at $> 4\times 10^5~\mathrm{K}$ to be detected by AIA.

\end{abstract}


\keywords{Sun: atmospheres-corona- filaments, prominences- UV radiation}



\section{Introduction}

Solar prominences are laboratories to investigate the physical conditions for maintaining the thermal and pressure stability of a partially ionized plasma embedded in a dynamic and fully ionized coronal environment. Prominences are cool (T $< 2 \times 10^4 ~\mathrm{K}$) and dense ($10^9 < n_e <10^{11}$ $\mathrm{cm^{-3}}$) structures  magnetically supported and insulated in the $> 10^6 ~\mathrm{K}$ corona. These structures may live for weeks before being ejected, or just disappear by fading out. 
Their interface with the corona, the Prominence - Corona  Transition  Region (PCTR), is thought to play a key role for this type of configuration. 
The PCTR is mainly made of  optically thin plasma emitting in the UV-EUV wavebands \citep{labrosse10}.

Recent spectroscopic and imaging UV-EUV investigations of prominences at the limb and filaments on the disk have revealed several new properties both of their cool core and their PCTR. The dark appearance of prominences and filaments in EUV images with respect to the nearby corona and the solar disk, respectively, can be attributed to  absorption and/or volume blocking \citep[e.g.][]{anzer05}. The first mechanism is the absorption of the background quiet-Sun emission by the dense prominence/filament core (mainly due to the photoionization of \ion{H}{0} and \ion{He}{0}; \cite{kucera98,heinzel01b, schmieder03}), the second one results from the presence of a volume of space 
having weak or absent prominence/filament emission at temperatures corresponding to  the respective wavelength bands \citep{schmieder04,schwartz04,schwartz06}. Several studies have addressed the disentangling of these two phenomena \citep[e.g.][]{delzanna03,  heinzel08}, necessary to diagnose the filament/prominence. For example, quantifying the absorption allows the determination of the column mass of prominences and the ionization degree of hydrogen and helium. 
The lack of emission at certain wavelengths can be interpreted as a too small, or 
even absent, hot prominence mass emitting at the corresponding TR and coronal temperatures. To establish the temperature at which this low mass limit is reached is of extreme importance for understanding PCTRs. This will, for example, impose constraints on the energy balance of the structure.

When the filament is observed in  UV-EUV images as a prominence on the limb, further information can be obtained. At coronal temperatures, quiescent prominences generally appear dark against the bright off-limb corona. More ambiguous is their appearance when observed in  transition region (TR)  emission, because they may not be seen at all, or they may be visible in absorption against the background corona. The latest generation of high spatial resolution images with high signal-noise ratio of SDO/AIA \citep{lemen12}, is revealing that quiescent prominences may also be weakly seen  in emission in the high-temperature TR, as shown by the 171 \ion{Fe}{9} images (see for example Figure \ref{fig_aia}). 

Spectroscopic data provide a complementary tool to study the thermal structure of prominences. The access to multiple spectral lines having  formation temperatures which span a wide temperature range provides the tool to derive the Differential Emission Measure (DEM). This allows us to quantify the amount of emitting plasma over a large range of temperatures along the line of sight, as  has been shown by \cite{parenti07} and \cite{gunar11} (see Figure \ref{fig_dem}). 


Inferring the Differential Emission Measure  in principle also allows us to establish the highest temperature emission in the prominence detectable by our instruments. In a real situation, however, the task is more complicated. Being optically thin, the hottest emissions of the prominence are integrated together with those of the foreground and background off-limb emissions along the same line of sight. Independent measurements of the off-limb quiet emission, free of all structures, would be necessary for disentangling these emissions.  These measurements, however,  are not always available. This was the case for the 1999 and 2004 prominence observations plotted in Figure \ref{fig_dem} and used for the present work: their inferred DEM increases between $ 5.2 <\log T < 6.1 $, as it does for a typical DEM derived from quiet off-limb observations. Under these conditions we are not able to say how much of this emission measure belongs to the prominence.

The present work is a first step toward answering this problem. 
We have mentioned that the AIA 171  band, which peaks in \ion{Fe}{9} 171.07 \AA ~at $\log T=5.8$, reveals weak  emission in prominences. 
This could mean either that the PCTR is hotter than thought before or that some cool emission dominates the band over the expected $\log T=5.8$ emission. To answer this question is the goal of the present work (see also \cite{heinzel12} for some preliminary results).

The investigation of cooler emission in the AIA bands has been the subject of a few studies, and the main result concerning the 171 band is  that  \ion{Fe}{9} remains the dominant component under a variety of different physical conditions \citep{odwyer10, delzanna11}. These studies used on-disk data and none of them included the case of  prominence emission. 
In this work, instead, we  use prominence data. Because a prominence consists mostly of cool plasma, our aim is to test whether the emission of any cool contributor can become dominant over that of \ion{Fe}{9} ion in the 171 AIA band. This investigation will also be extended to other AIA bands.

The plan of the paper is  the following: Sect. \ref{aia_em} presents a case of a prominence in emission as observed by AIA;  we then build a synthetic AIA  prominence spectra in Sect. \ref{synt_spect} and discuss the results and conclusions in Sect.  \ref{inter}.

\section{A case study for AIA}

\label{aia_em}

\begin{figure*}[th]
   \centering
\includegraphics[scale=0.5]{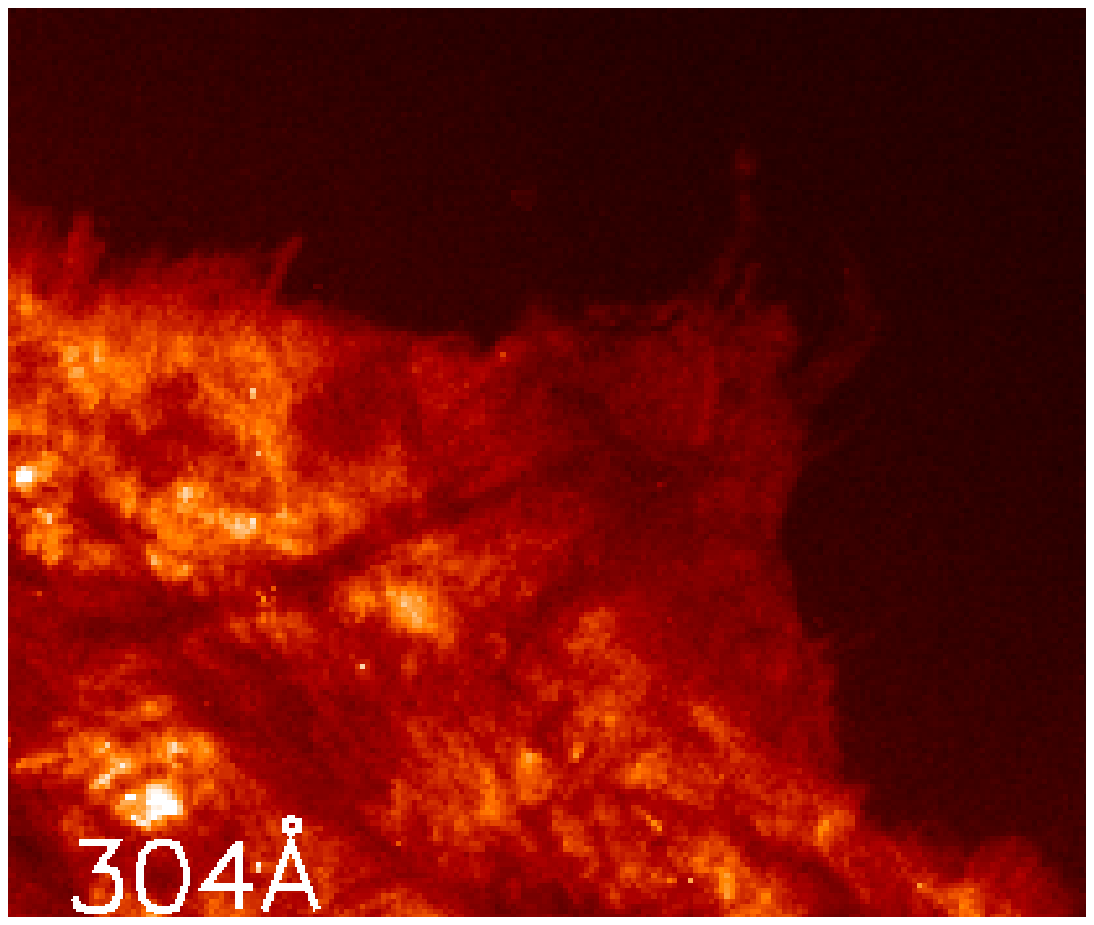}
\includegraphics[scale=0.5]{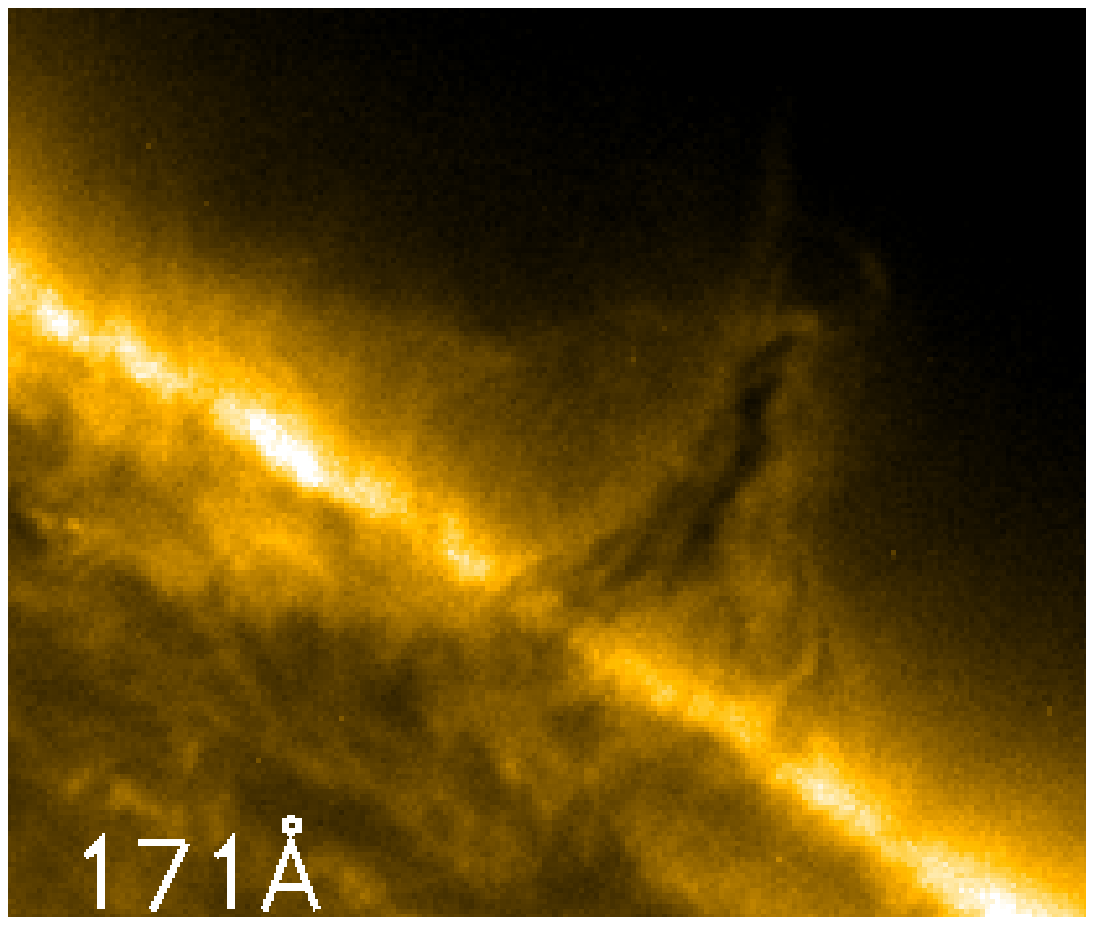}\\
\includegraphics[scale=0.5]{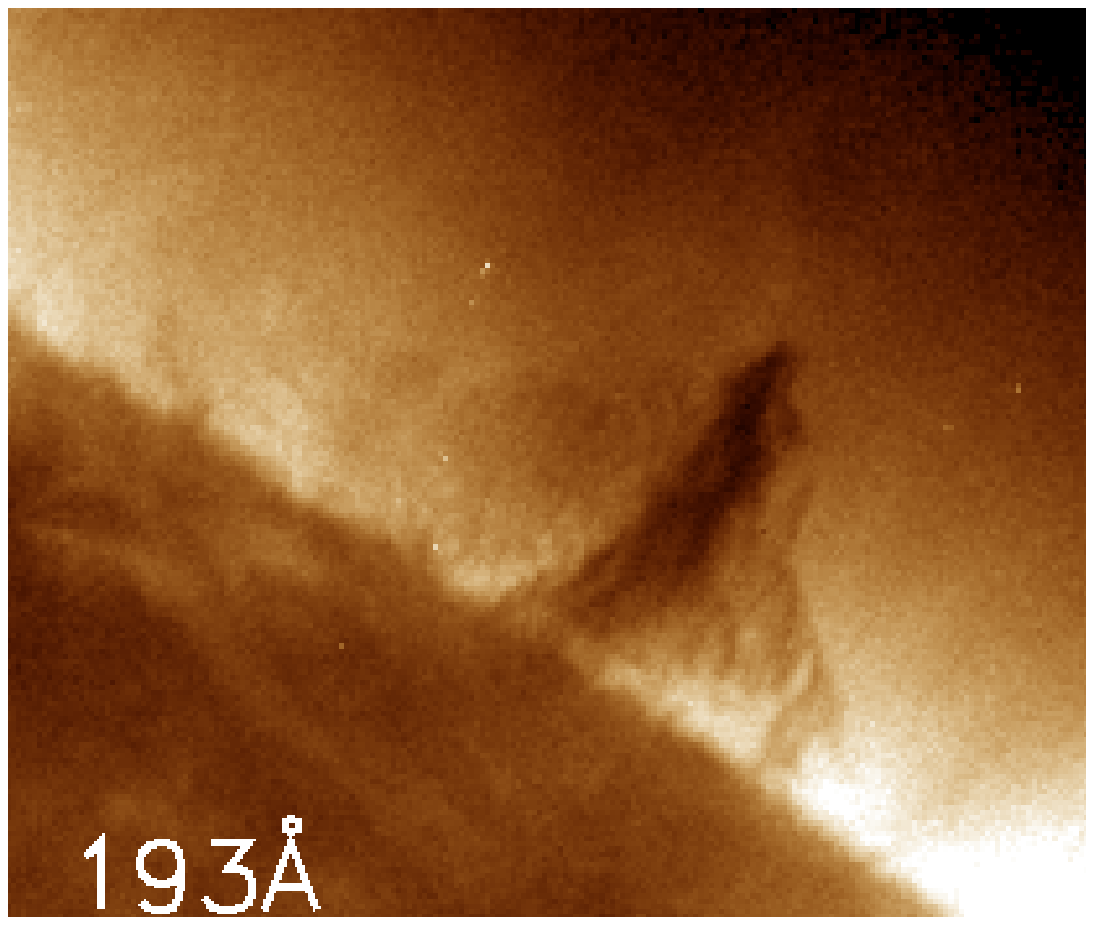}
\includegraphics[scale=0.5]{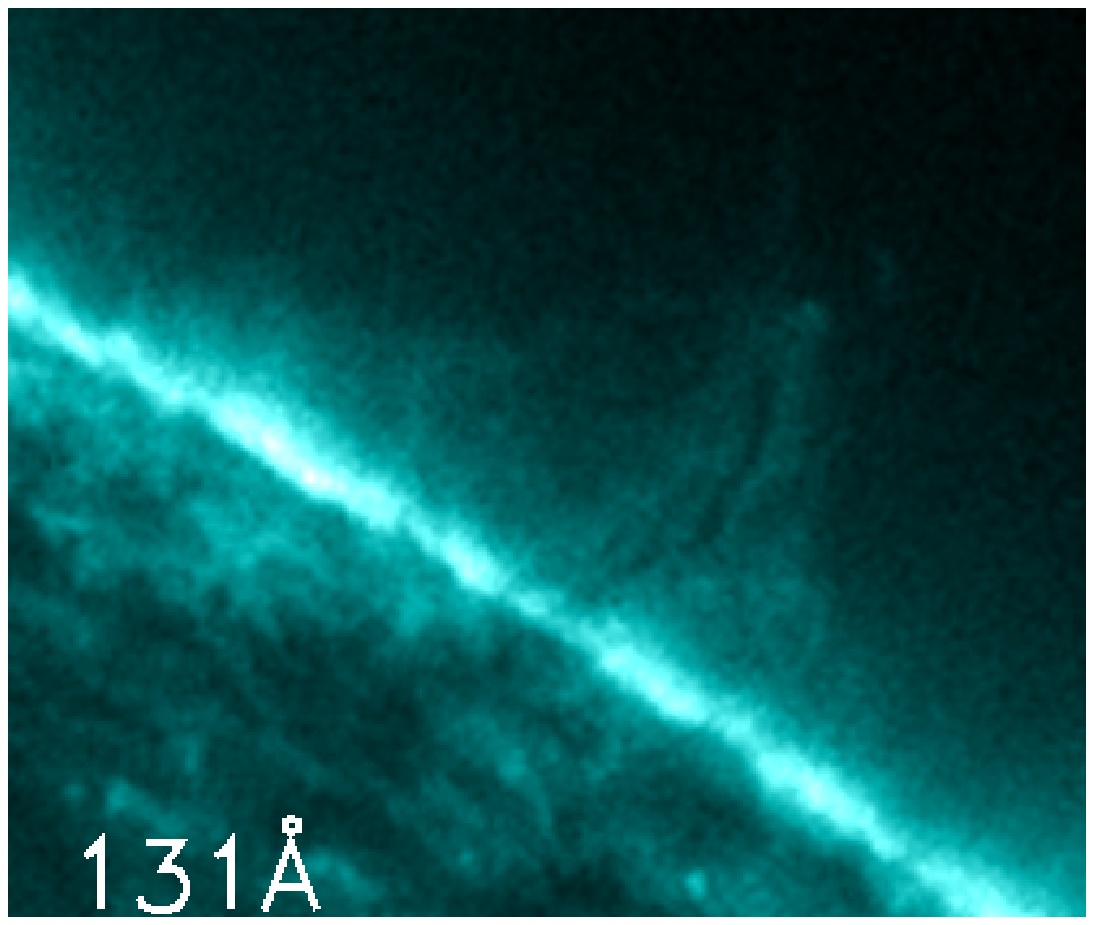}
   \caption{SDO/AIA images of a quiescent prominence observed at 15:38 UT on the limb on 22 June 2010 over a $124" \times 108"$ field of view. From top left to bottom right: single exposure for 304 band ($\log T=4.7$), 171 band ($\log T = 5.8$),  193 band ($\log T = 6.1$) and 27 sec exposure for 131 ~band ($\log T = 5.8$ and 7). } 
\label{fig_aia}
    \end{figure*}

\begin{figure}
   \centering
\includegraphics[scale=0.5]{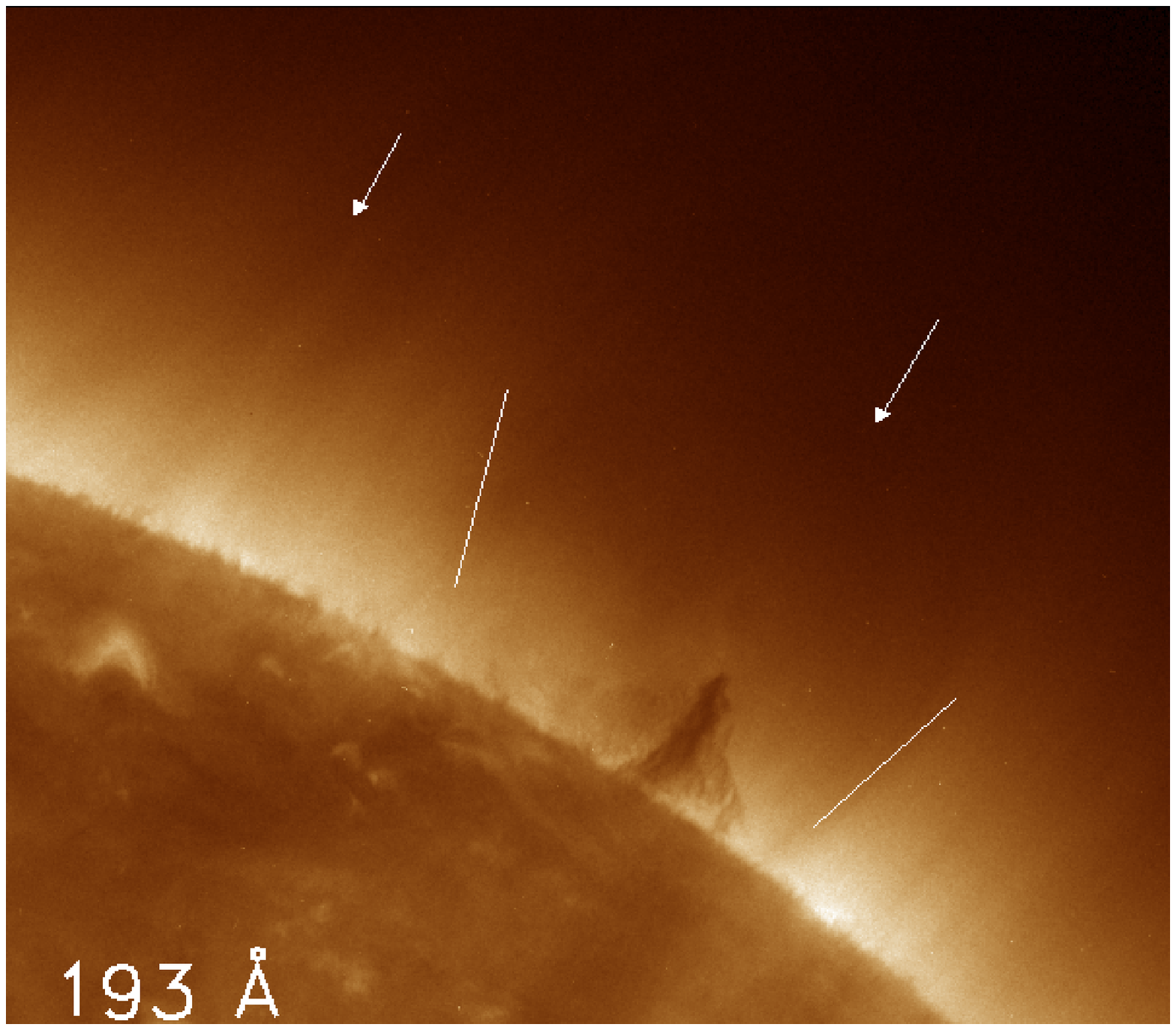}
\caption{SDO/AIA image in the 193 band showing a large field of view. Here we can detect the  coronal cavity (marked by the solid white segments). The arrows mark the location of the radial cuts used to plot the profiles in Figure \ref{fig_aia_prof}.}
\label{fig_aia_193}
    \end{figure}

Figure \ref{fig_aia} shows images of a NW prominence observed by SDO/AIA on June 22, 2010. A fine scale dynamic study of this prominence was done by \cite{berger11}.
The structure is  visible in emission in the transition-region dominated 304  band (top-left, 2.9 sec exposure, $\log T$ =4.7) and its dense spine is  seen in absorption in the coronal emission dominated 193  band (left-bottom, 2 sec exposure, $\log T$ = 6.1-7.3). Depletion of the intensity with respect to the background corona, even if of lower amplitude then in the spine, is seen in the 193 band in the volume occupied by the prominence. 
The transition region 171 band (2 sec exposure) on the top-right 
panel shows the intensity variation inside the prominence which resembles  a combination of the 304 and 193 bands:  the spine is dark and
a faint emission is visible in the rest of the prominence area. 
The bottom-right image shows the 131 band. This band is dominated by flaring lines emitting in the range $\log T$ = 7-7.2, but it also contains transition region lines, especially \ion{Fe}{8}. 
To obtain a low noise image we integrated the data for about 27 secs (nine images). With this integration time we are able to see the prominence dark spine and a weak emission in the rest of prominence. The coronal cavity is also visible in the 193 band when we image a more extended field of the  view, as shown in Figure \ref{fig_aia_193}.

\begin{figure*}[ht]
   \centering
\includegraphics[scale=0.4]{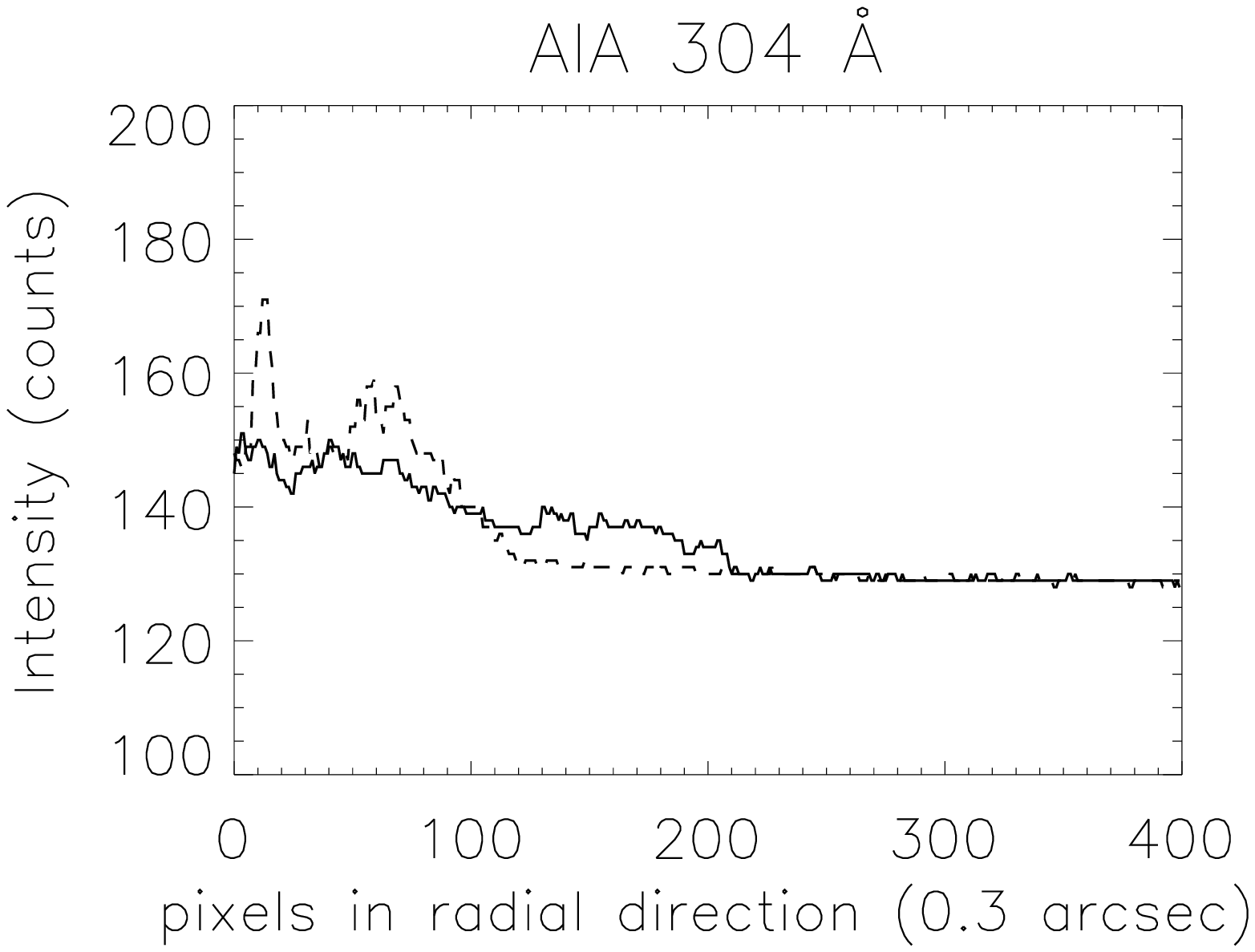}
\includegraphics[scale=0.4]{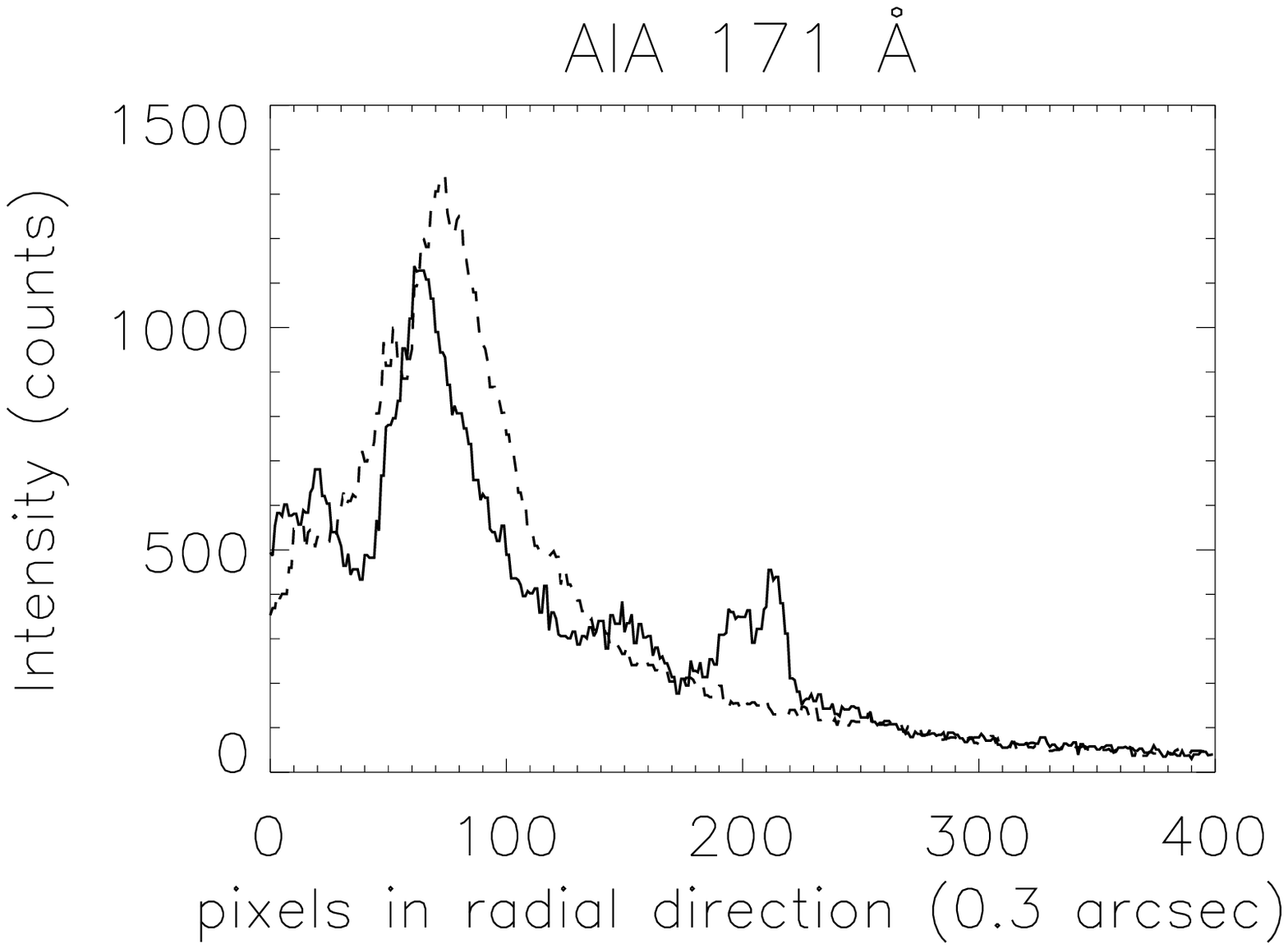}\\
\includegraphics[scale=0.4]{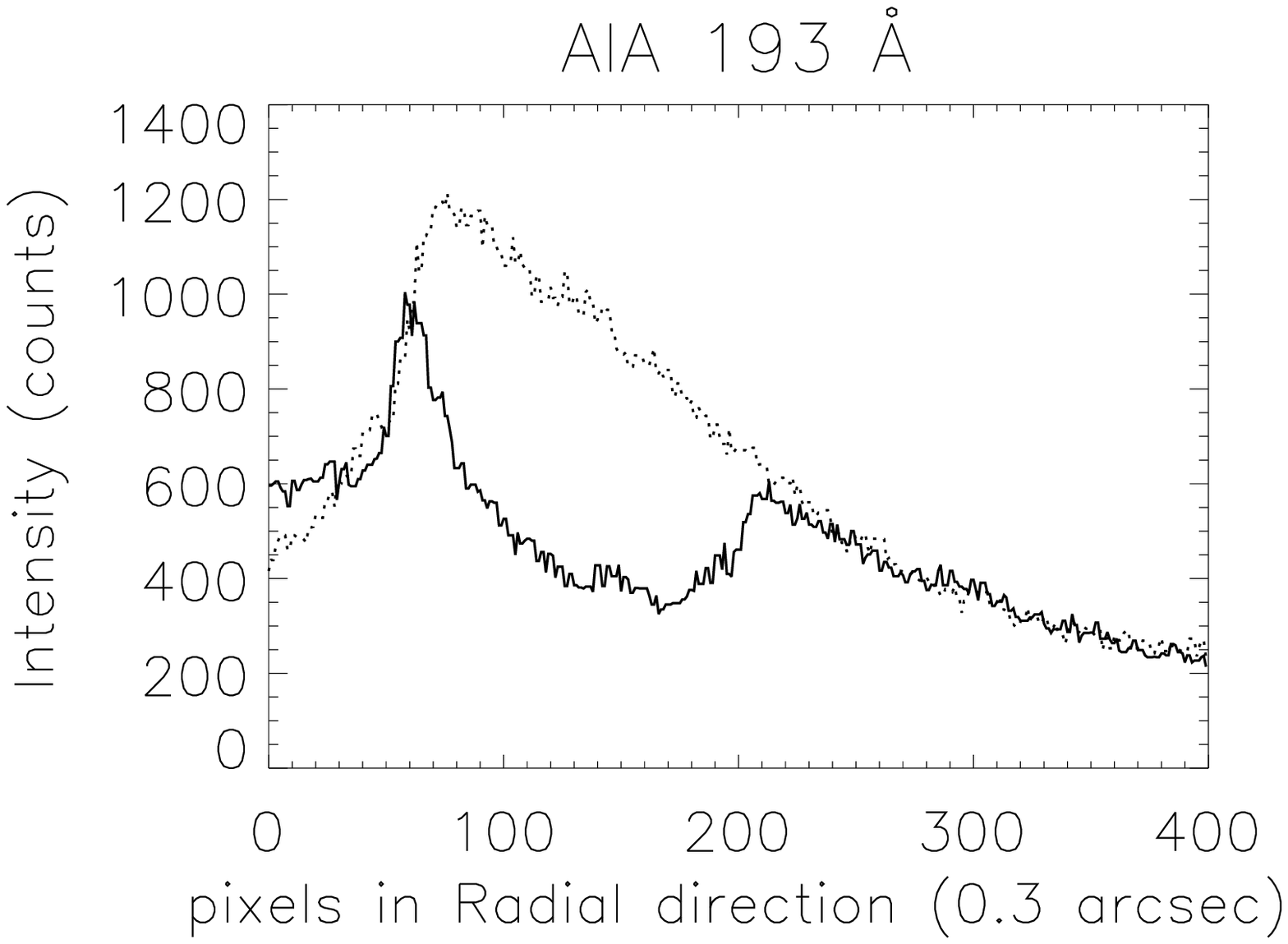}
\includegraphics[scale=0.4]{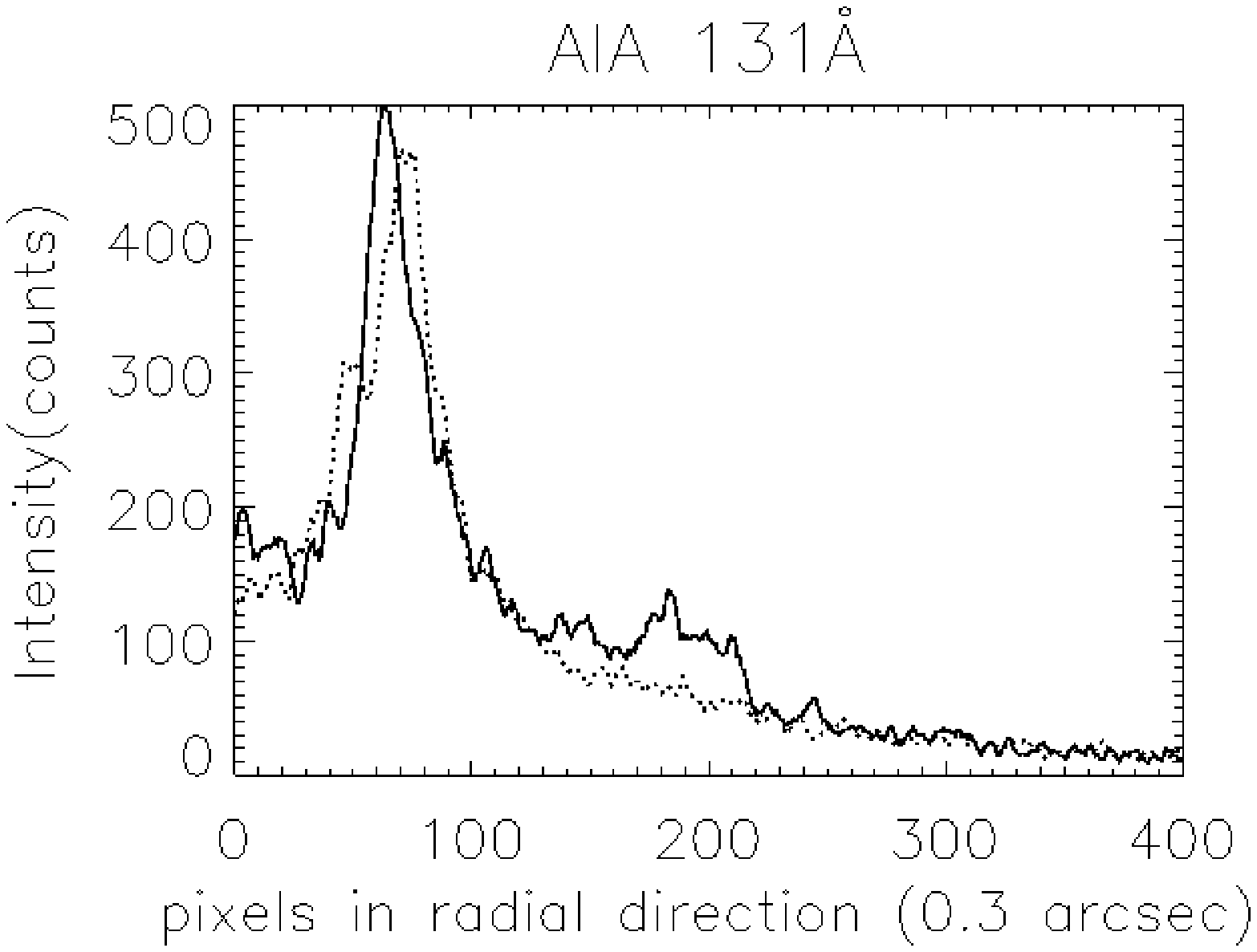}
\caption{AIA radial intensity profiles in the 304, 171 and 131 bands. The solid line is for the prominence and the dashed line for the quiet corona. Emission above the level of the background corona is clearly seen inside the prominence in the 171 and 131  bands.}
\label{fig_aia_prof}
    \end{figure*}

Figure \ref{fig_aia_prof} quantifies this emission/absorption in each band by comparing the off-limb quiet-corona radial profiles with those inside the prominence. The location of these radial cuts are marked by the arrows in Figure \ref{fig_aia_193}. 
In the plots the solid line refers to the prominence and the dashed-line to the corona. In the 193   band the prominence is mostly  in absorption. This is caused mainly by photoionization of the neutral $\mathrm{H}$, $\mathrm{He}$ and $\mathrm{He^+}$ by the coronal  radiation passing through the prominence at this wavelength. Considering the proximity of the 171 to the 193 bands, we  expect a similar absorption behavior. Partial absorption is still visible inside the dense spine at 171, however starting from pixel 140 the prominence emission overcomes the quiet-off limb corona. Surprisingly, the 131 prominence profile follows a similar behavior.


Correlated with the 304  band prominence profile, the intensity above the coronal background shown by the 171  and 131  bands inside the prominence is not of a coronal nature. The origin of this emission is the subject of the following analysis.

\section{Synthetic spectra}
\label{synt_spect}

\begin{figure}[th]
   \centering
\includegraphics[scale=0.4]{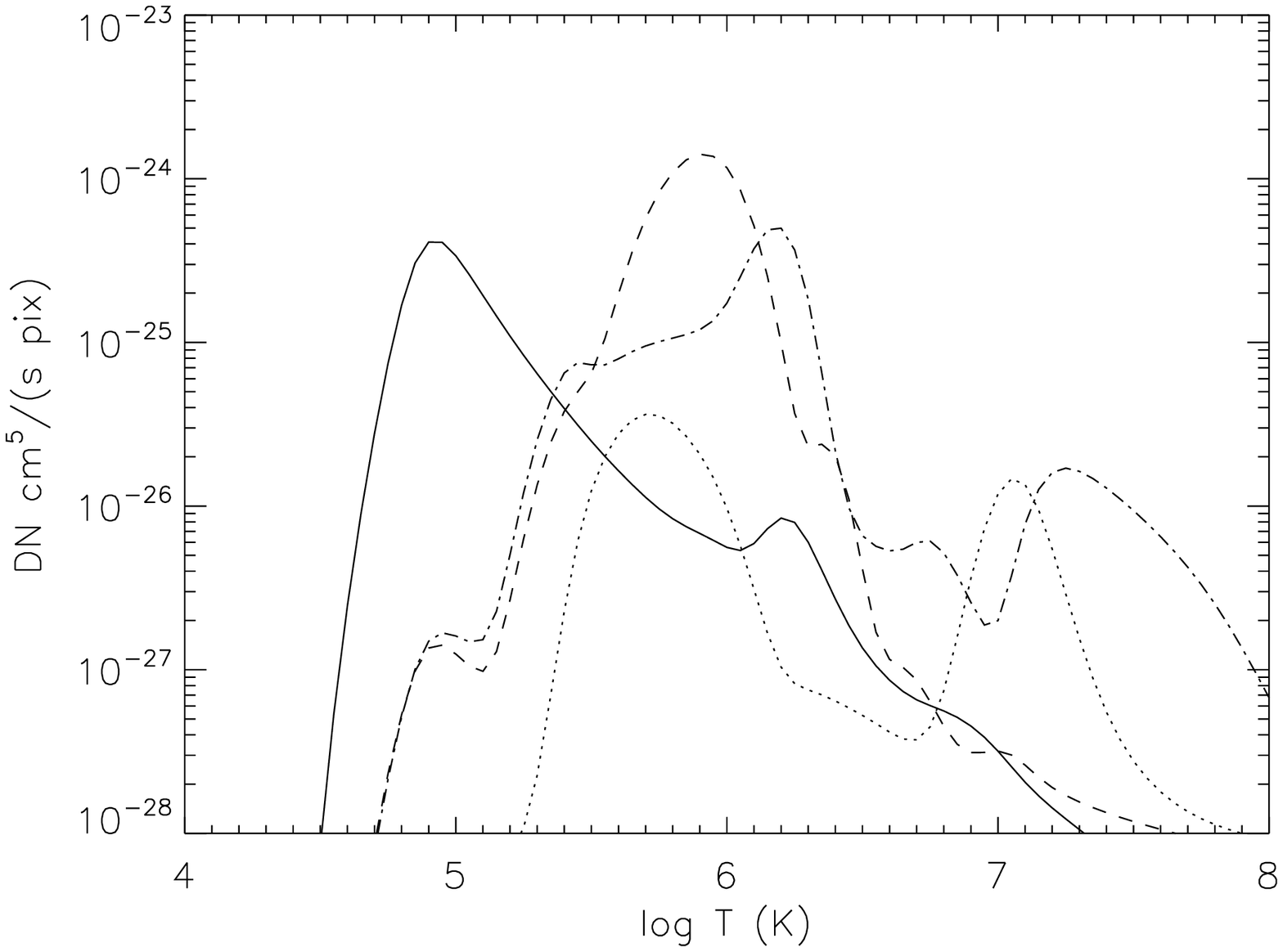}
   \caption{Response function of AIA 131 (dotted line), 171 (dashed line), 193 (dash-dotted line) and 304 (solid line)  bands as function of  the temperature.}
\label{aia_tresp}
    \end{figure}

In order to interpret the emission in the AIA bands, we performed  a series of simulations  producing prominence synthetic spectra in these bands. Our aim is to estimate the contribution of cooler lines to the total measured intensity. The following analysis is done on three representative AIA bands: 171 for the  transition region emission, 193 for the coronal emission and 131 for the transition region and flaring emission. Their response curves as function of temperature are plotted in Figure \ref{aia_tresp} together with that of the 304. 
This Figure shows  multiple peaks in each filter, suggesting that the measured intensity may not be easily interpreted. Nevertheless, the intrinsically brightest lines in each band, together with their temperature at the peak formation, are:  \ion{Fe}{8} 131 \AA ~and \ion{Fe}{21} 128.75 \AA~($\log T $= 5.6 and 7.05);  \ion{Fe}{9} 171.0733 \AA ~($\log T$ = 5.8);  \ion{Fe}{12} 193.51 \AA~and \ion{Fe}{24} 192.03 \AA~($\log T $=6.2 and 7.25).

\begin{figure}[th]
   \centering
\includegraphics[scale=0.35, angle=90]{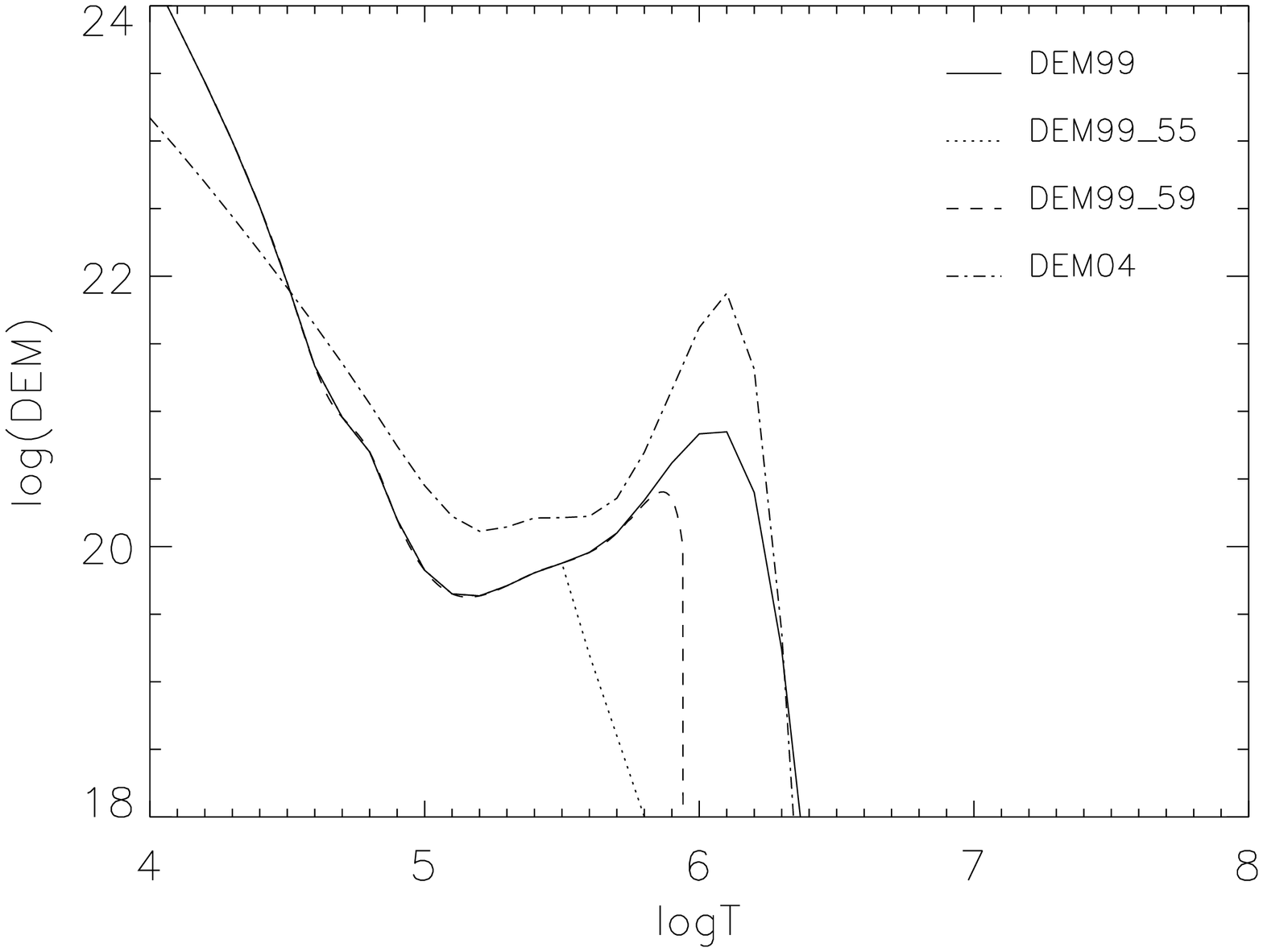}
   \caption{DEM profiles used in this work: DEM99, DEM99$\_55$, DEM99$\_59$, DEM04.}
\label{fig_dem}
    \end{figure}

To build the spectra we used two prominence DEMs obtained using SOHO/SUMER data. The first DEM, called DEM99 in  Figure \ref{fig_dem} (solid line, \cite{parenti07}), was inferred  using a sub-set of the line  intensities (about 50) listed in the spectral atlas of the prominence observed in 1999 by \cite{parenti04, parenti05a}. The second DEM, called DEM04 in  Figure \ref{fig_dem} (dash-dotted line), was inferred by \cite{gunar11} from a prominence observed in 2004. These DEMs were obtained using optically thin lines non affected by continuum absorption, contrary to the spectra contributing to several of the AIA measured intensities. We will keep this in mind for the interpretation of our results.

DEM04 and DEM99 have different gradients at low PCTR temperatures ($< 10^5 ~\mathrm{K}$), but we are aware  that   these differences are due in part to a different value for the lower temperature constraining this part of the  DEM. 
This difference may imply different contributions of cool lines to AIA bands.  For higher  temperatures DEM04 runs almost parallel to DEM99, while it has a higher coronal contribution. For  these two prominences it was impossible to subtract the  background and foreground emission, so that the real contribution of the PCTR emission to these DEMs is unknown. However, for the purpose of our simulations, we will assume these DEMs as fully due to prominence emission. 
To study also the case of prominences with limited or no coronal emission (or background subtracted emission), we modified DEM99 by imposing very low DEM values above $\log T =5.9$ (DEM99$\_59$, dashed line) and above  $\log T =5.5$ (DEM99$\_55$, dotted-line). Both of these DEMs are plotted in Figure \ref{fig_dem}.

\begin{figure*}
   \centering
  \includegraphics[scale=0.4]{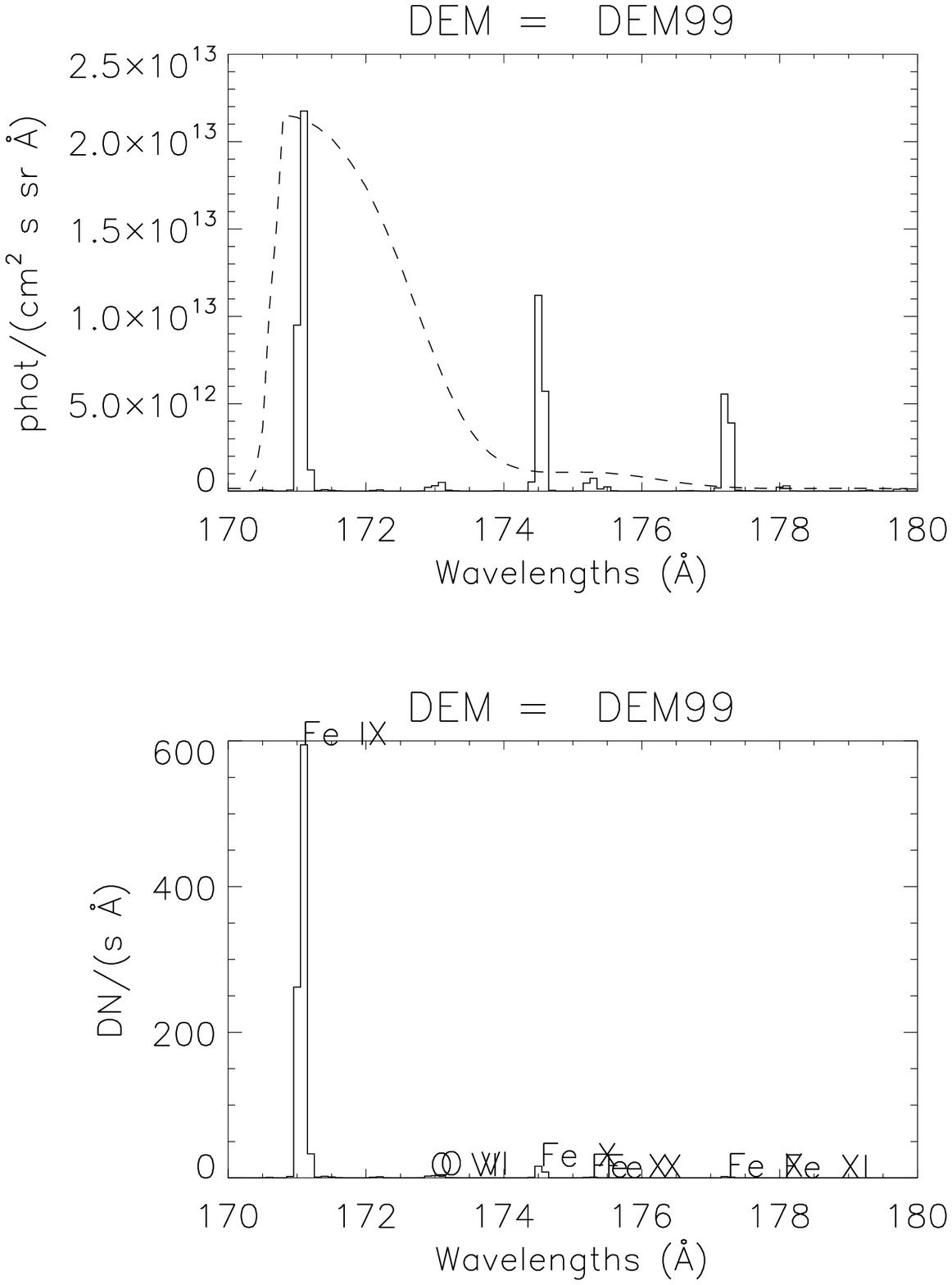}
\includegraphics[scale=0.4]{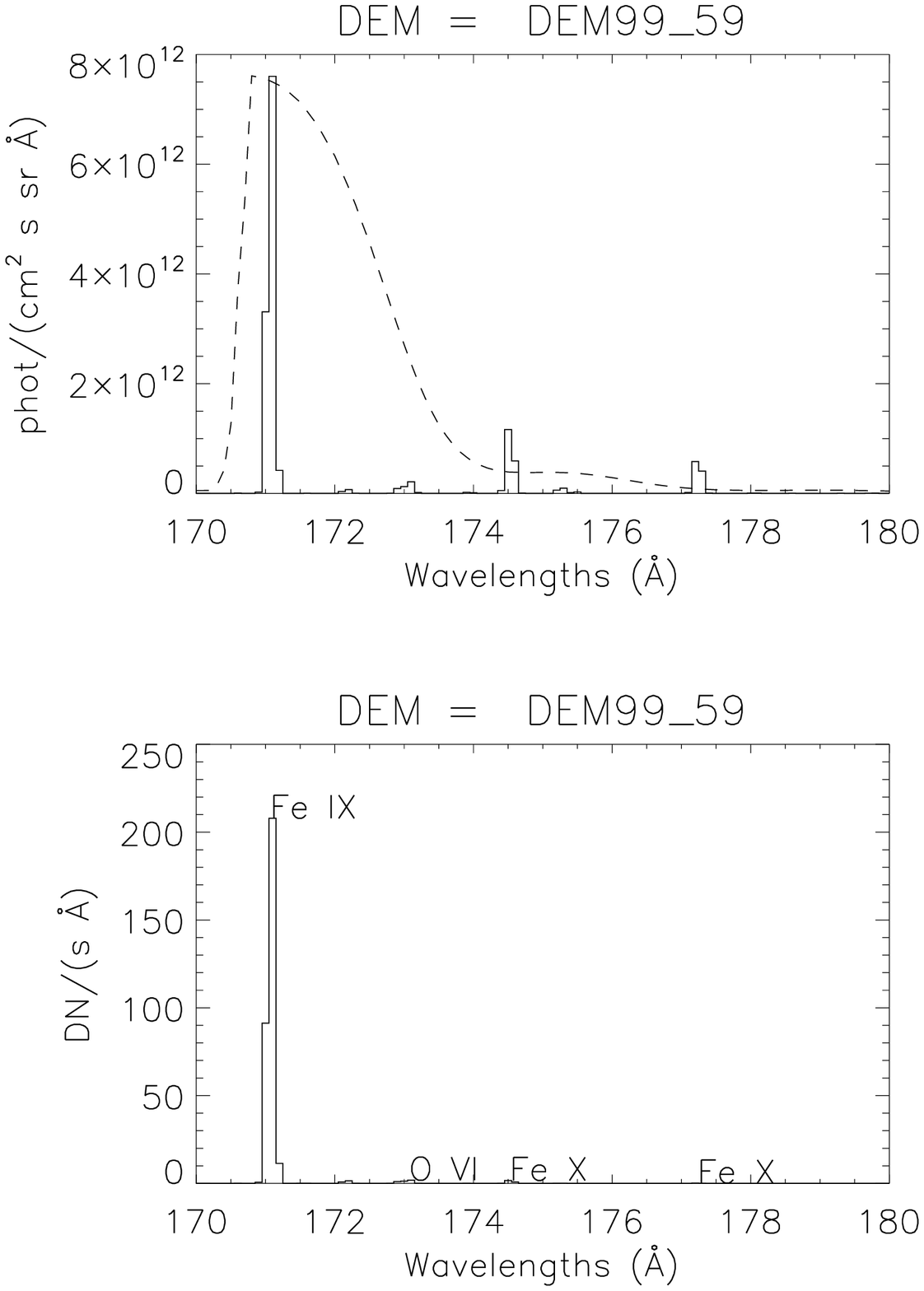}\\
\vspace{0.5cm}
   \includegraphics[scale=0.4]{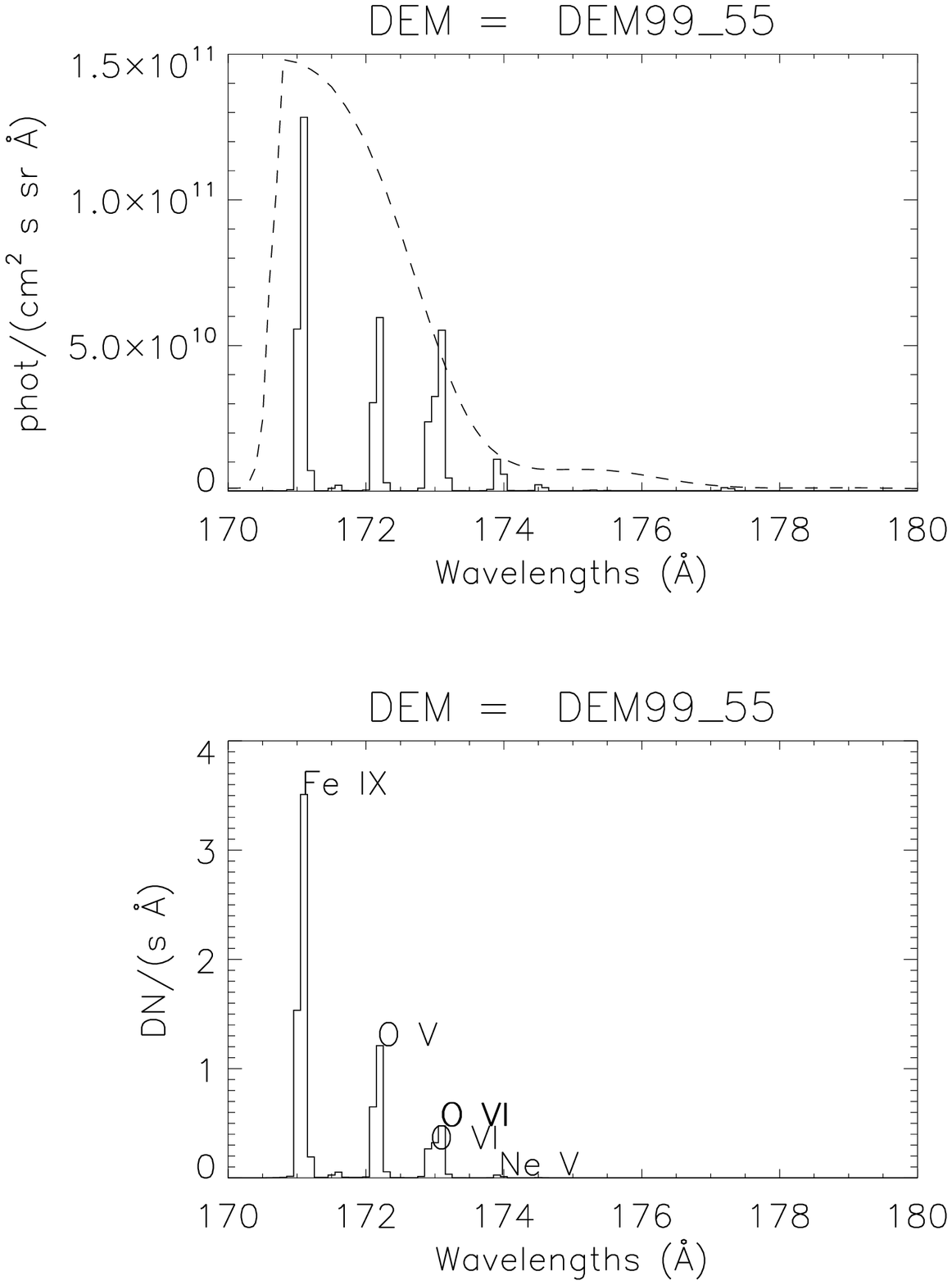} 
\includegraphics[scale=0.4]{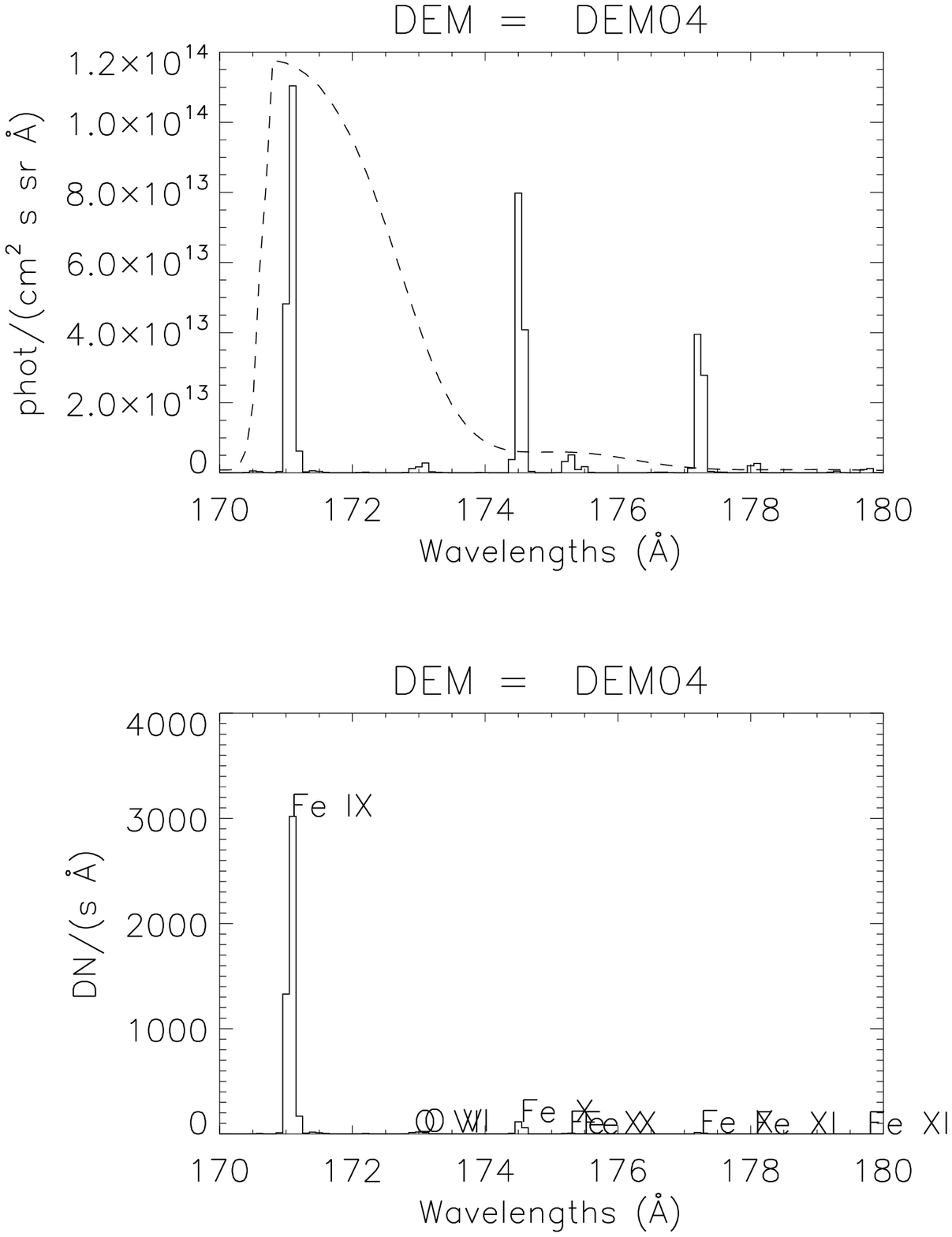}
   \caption{Synthetic CHIANTI spectrum contributing to the AIA 171  band derived using the different prominence DEMs plotted in Figure \ref{fig_dem}. For each DEM the top panel shows the prominence spectrum with superimposed the effective area of the AIA filter.  The bottom panel shows the spectrum predicted to contribute to the  AIA bands  obtained by the product of the spectrum shown in the top panel  with effective area of the instrument.}
\label{fig_spectra}
    \end{figure*}

Rows 1 and 3 of Figure \ref{fig_spectra} show the synthetic spectra in the 171 band of the brightest lines.  These were calculated using the CHIANTI v7 atomic database \citep{dere97, landi12}, assuming \cite{mazzotta98} ionization equilibrium, photospheric abundances and a  constant pressure of 0.1 $\mathrm{dyne ~cm^{-2}}$.  These values were used to infer the DEMs from the SUMER data. 
Superimposed in the spectrum is the effective area of the filter, as available from the AIA software on $SolarSoft$. 

Rows 2 and 4 of Figure \ref{fig_spectra} are the resulting spectra after 
multiplication with the effective area.
These plots show that  the high-temperature transition region  \ion{Fe}{9} 171.0733 \AA~line dominates the band in any circumstance, even though the DEMs shown in Figure \ref{fig_dem} have a different profile, particularly at  low TR temperatures. This can be  explained by  the lack of intrinsically bright cool lines inside the band. 
Table \ref{tab_int} quantifies the contributions of the brightest lines in the band for each of the four DEMs, while Table \ref{tab_counts} reports the simulated total counts/s. 
Table  \ref{tab_int} shows that the cooler lines (i.e. from O ions) have an important weight (about $30\%$) on the bands intensity only when we use DEM$\_55$. However, the DNs in the band decrease considerably and the band would become dominated by noise for standard exposure times (2 secs). 
In all the cases shown here the continuum emission (free-free, free-bound and two photon continua are included in the CHIANTI) contributes significantly only in 131 band (i.e. a relative contribution).

\begin{figure*}
   \centering
  \includegraphics[scale=0.4]{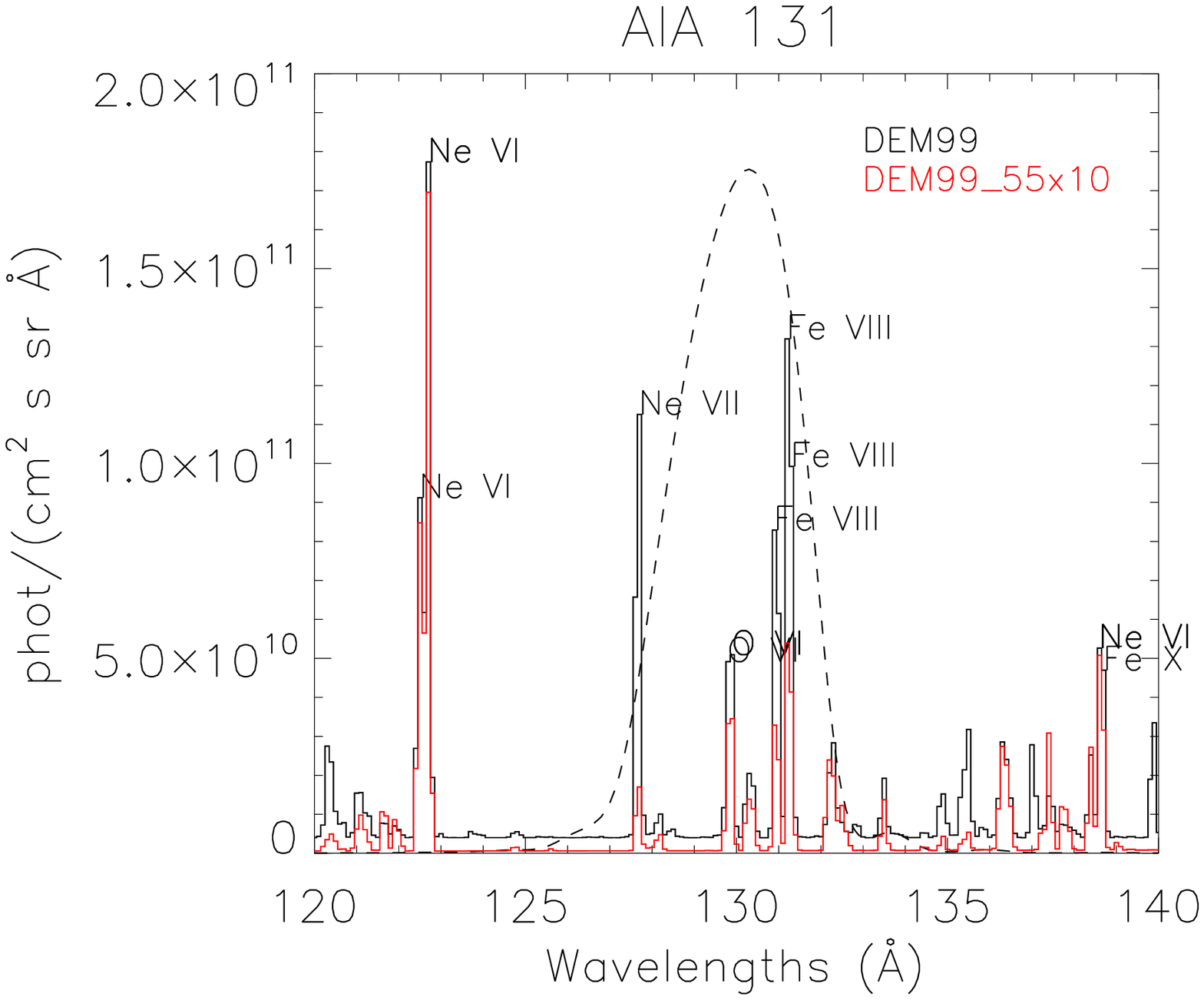}
  \includegraphics[scale=0.4]{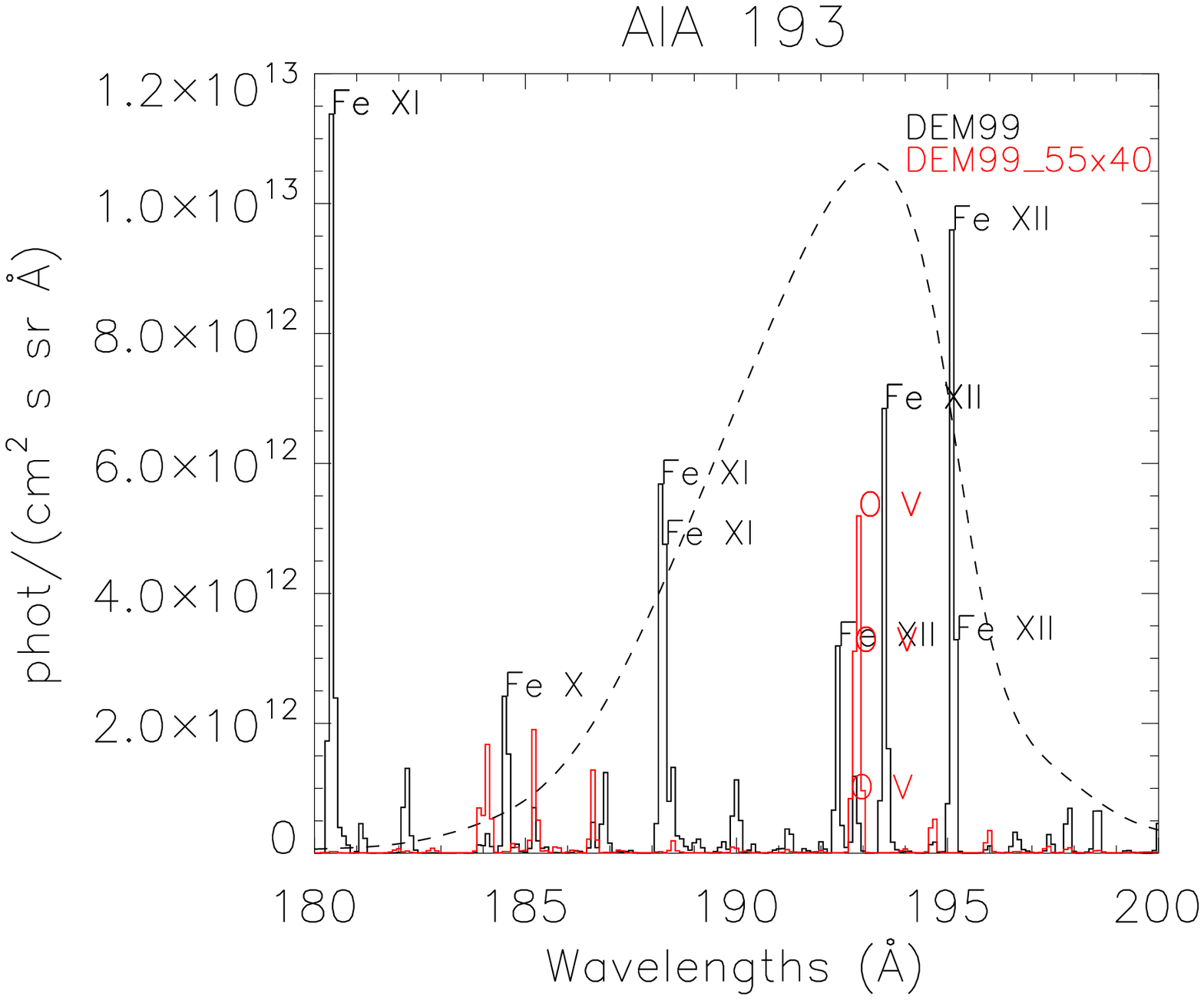}\\
  \includegraphics[scale=0.4]{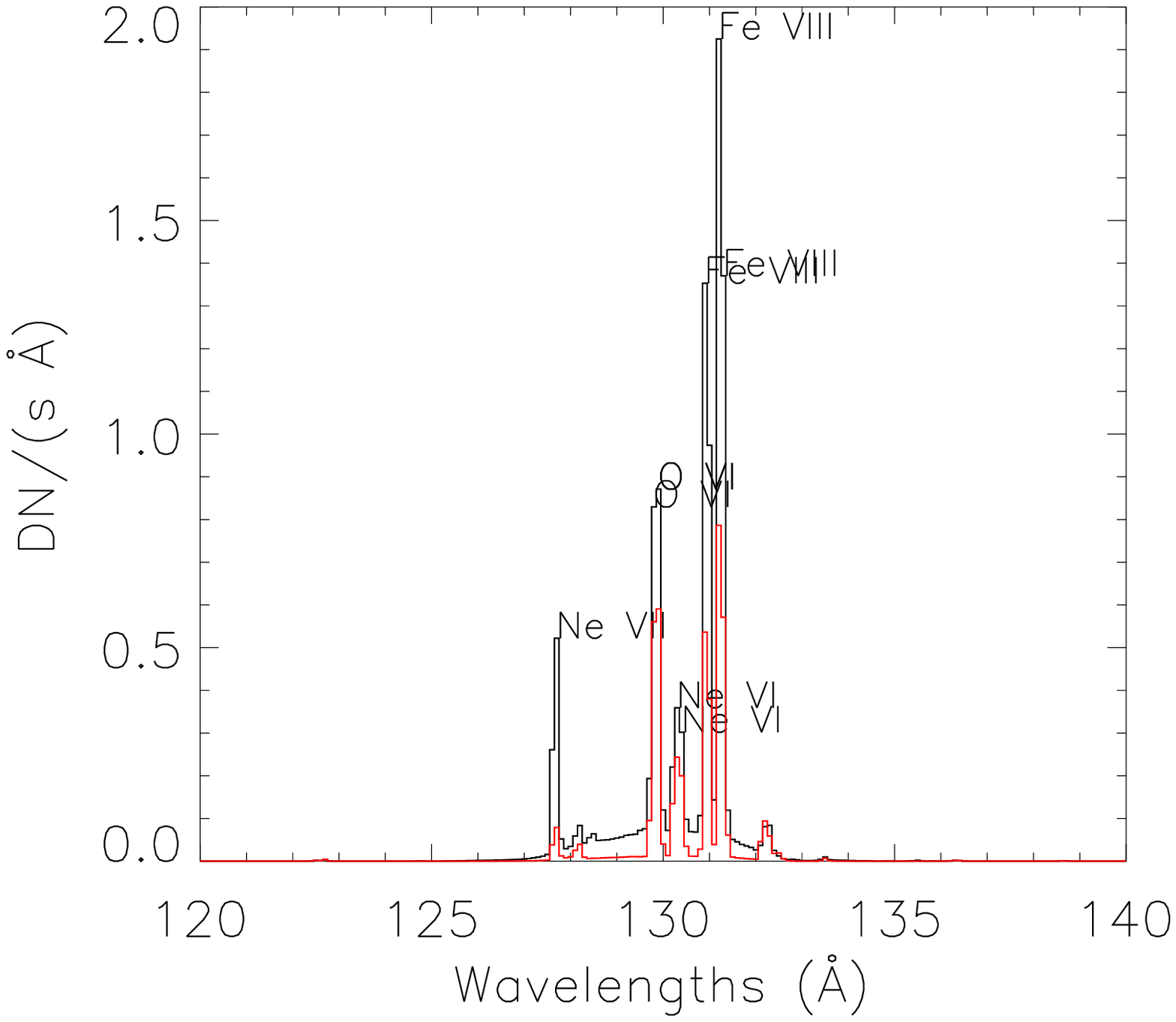}
  \includegraphics[scale=0.4]{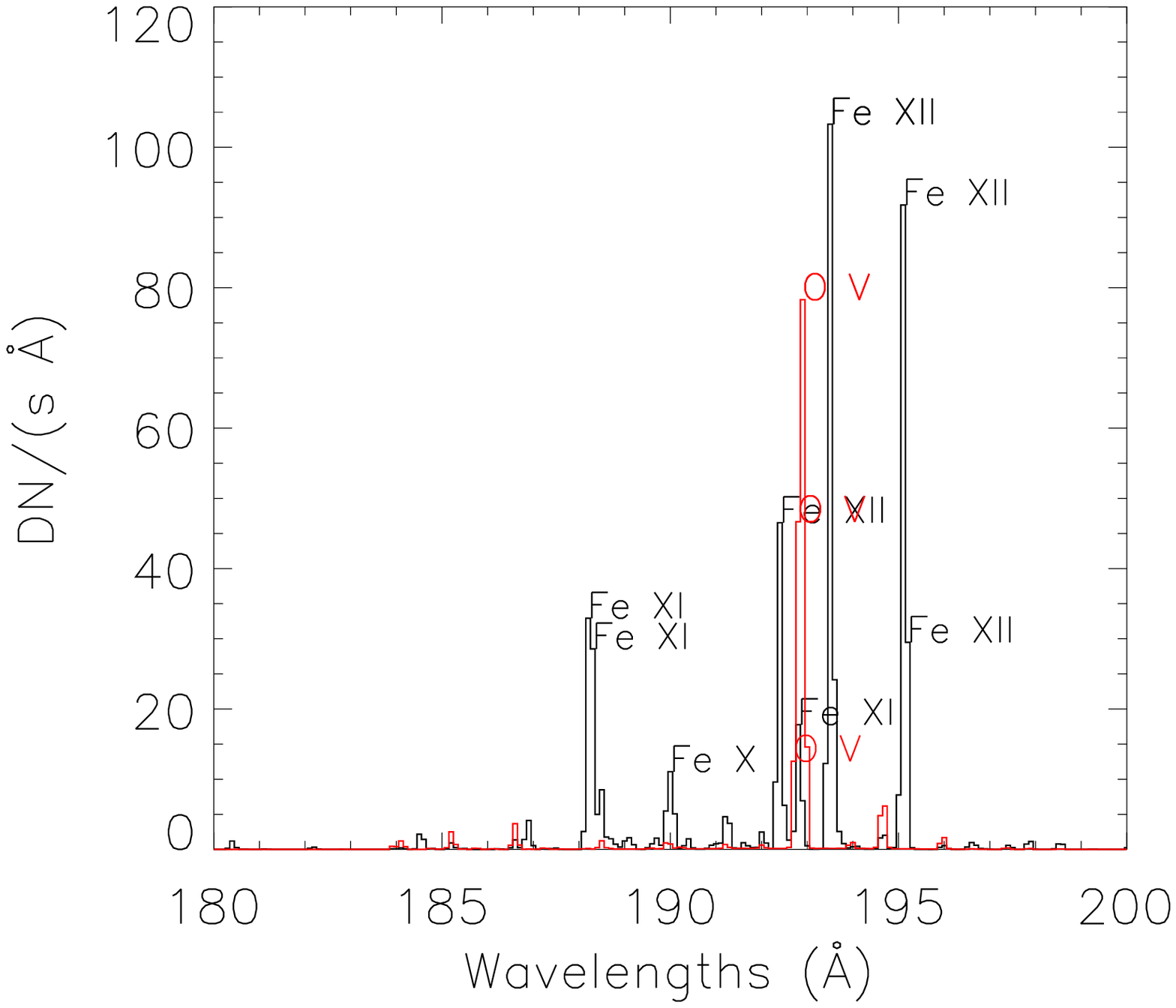}
 \caption{ Synthetic prominence spectrum in the AIA 131 (left) and 193 (right). Top: prominence spectra with superimposed the effective area (dashed line). Bottom: the resulting  spectrum predicted to contribute the bands  after multiplication with the effective area of the filter. Each color refers to a different DEM. The DEM99$\_$55 spectrum, in red, has been multiplied by a factor 10.}
\label{fig_sp_hot}
    \end{figure*}

On the basis of these results we built synthetic spectra also for the 193 and 131  bands. For their calculations we used DEM99 and DEM99$\_55$ as representative of the two extreme cases, which are a prominence surrounded or not surrounded by a coronal environment. The results are plotted in Figure \ref{fig_sp_hot} and the contribution of the main lines is listed in Table \ref{tab_int}.
 
Without any flaring component in these DEMs, the 131 intensity is dominated by the emission at temperatures $\log T = 5.6-5.7$ (between about 50-60$\%$) with \ion{Fe}{8} being the main contributor. The continuum emission also contributes.  
However, looking at the simulated intensity in the band (bottom panel of Figure \ref{fig_sp_hot} and Table \ref{tab_counts}), it accounts only for a fraction of a DN/s, and considering that the typical exposure time is  3 sec, this emission is hidden in the noise. We integrated over several images to obtain the profile of Figure \ref{fig_aia_prof}.


The situation is  different for the 193 band. When the DEM99 is used, the measured intensity will be mainly due to \ion{Fe}{11} and \ion{Fe}{12}, that is from plasma above $\log T = 6.1$. On the contrary, when the DEM99$\_55$ is used, the \ion{O}{5} will be the dominant emitter, while there will be no sign of coronal intensity. However, in the latter case, for the typical exposure time of 3 sec only noise will be detected in the band. 

 Unfortunately we cannot extract more information from these simulations. Table \ref{tab_counts} tells us that the simulated 193 intensity is always lower than the 171, while Figure \ref{fig_aia_prof} shows  the opposite. One reason is the difference in the coronal foreground and background level for the three prominences observed in 2010, 2004 and 1999. 
  We can say, however, that because we have counts in  exposure time of only a few seconds in the 193 and 171 bands for the prominence of Figure \ref{fig_aia}, its DEM is somewhere in between DEM99 and DEM99$\_$59, but certainly not DEM$\_$55.

\begin{deluxetable}{rrrrrrrr}
\tablecolumns{8}
\tabletypesize{\footnotesize}
      \tablecaption{Major relative contributors to the AIA bands for different  DEMs. \label{tab_int}}
        
\tablehead{
          \colhead{Filter} & \colhead{Line} & \colhead{$\lambda$} & \colhead{T} & \colhead{DEM99} & \colhead{DEM99$\_59$ } & \colhead{DEM99$\_55 $} & \colhead{DEM04} \\
  \colhead{}     & \colhead{} &   \colhead{\AA}     & \colhead{$\mathrm{K}$} &
   \colhead{$\%$}     & \colhead{$\%$}     &   \colhead{$\%$}     & \colhead{$\%$} \\ 
}
\startdata
              171  &  \ion{Fe}{9} & 171.0733 &  5.8 & 94 & 96 &  61 &  92  \\
						&\ion{Ni}{14}&  171.3703 & 6.3 & 0.3&  -   &  - & 0.5\\
						&\ion{O} {5}  &  172.1690 & 5.4 & 0.3 & 0.7 & 22 &  -  \\  
						&\ion{O}{6}  & 172.9357  & 5.5  & 0.4 & 0.5 & 5 &  0.5   \\
						&\ion{O}{6}  & 173.0798  & 5.5  & 0.6 & 0.7 & 7 &   0.6  \\
						&\ion{O}{6 } & 173.0951  & 5.5   &  - &  -  & 0.8 &   -   \\
						&\ion{Ne }{5 } & 173.9320 & 5.5 & -  &  -  & 0.5 &   -   \\
						&\ion{Fe}{10}  & 174.5310  & 6.0  & 2.7 & 0.8 &  - &  3.6    \\
						&\ion{Fe}{10} &  177.240 &  6.0  & 0.3 & -   &  - &    0.4  \\
                  & Continuum   &           &  &   0.3 & 0.2 & 0.1 &   0.4  \\
                \hline
              131 & \ion{Ne}{7}  & 127.666 &  5.7  & 7  & & 2.3  & \\
						&\ion{O}{6} &  129.7814   & 5.45 & 4.9 &  &  8.1  &    \\
 						&\ion{O}{6} &  129.87    & 5.45   & 10 &  &   17 & \\
						& \ion{Ne}{7}  & 130.263 &  5.7  & 3  & &  7 & \\
						& \ion{Ne}{7}  & 130.394 &  5.7  & 2  & & 4  & \\
						&	\ion{Fe}{8} & 130.9412 & 5.6 & 19 & & 18  &\\
 					    & \ion{Fe}{8} & 131.2402 &  5.6  & 28  &  & 27 & \\
                   & Continuum   &           &      &  27 &  & 9   &  \\
               \hline
                193 & \ion{Fe}{11} & 188.2165 & 6.15  & 7.8   &  & - &  \\
 						&  \ion{Fe}{11}  & 188.29939 &  6.15 & 4.8  &  & - &  \\
 						& \ion{Fe}{12}  & 192.394 &  6.2  & 11.7  &  & - &  \\
						&\ion{O}{5}       & 192.8  &  5.4  &   -  & - & 27.3 & \\
						& \ion{Fe}{11}  & 192.81371& 6.15 & 4.2 &  & - &  \\
                   &\ion{O}{5}     & 192.904 &  5.4  &   -  & - & 44.6 & \\
						& \ion{Fe}{12}  & 193.509 &  6.2  &  25.7 &  &-  &  \\
 						 & \ion{Fe}{8} & 194.661&  5.6  & -     &   & 5.2 &\\
 						& \ion{Fe}{12}  & 195.12  &  6.2  &  24.3 &  &-  &  \\
						 & Continuum   &      &      &  0.8 &  & 5  &  \\
					
\enddata
\tablecomments{The first two columns identify the line by the ion and reference wavelength. The temperature corresponds to that of the maximum of ionization fraction. The contribution of each line is given in percentage with respect to the total emission in the band.}
\end{deluxetable}

 \begin{figure}[th]
   \centering
\includegraphics[scale=0.3, angle=90]{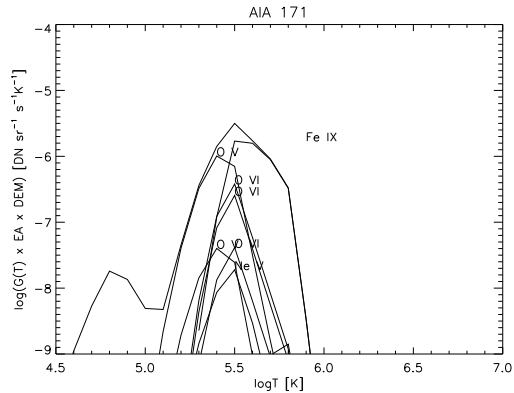}
   \caption{Contribution functions of the most intense lines in the 171 band, multiplied by the effective area of the instrument and the DEM$\_55.$}
\label{fig_gts}
    \end{figure}

\begin{deluxetable}{rrrrr}
\tablecolumns{5}
\tablewidth{0pt}
      \tablecaption{Simulated total counts in the AIA bands (DN/s) \label{tab_counts}}
        
\tablehead{
\colhead{Filter} & \colhead{DEM99} & \colhead{DEM99$\_59$ } & \colhead{DEM99$\_55 $} & \colhead{DEM04} \\
}
\startdata
              171  &  95 &  32 &   $<1$ & 490\\
               131 &  1   &$<1$ & $<1$ & 6\\
               193 & 53  & 3  &  $<1$    &    460\\
\enddata
\end{deluxetable}

\section{Discussion}
\label{inter}

The appearance of a prominence against the background corona changes as function of the density, temperature and wavelengths. The main factors involved are:  the absorption of the coronal emission by the  hydrogen and helium resonance continua,  the change of  the  scale-height of the coronal-line emissivity depending on the temperature, and the  emissivity of the structure itself. We now discuss these three aspects.

Being close in wavelengths, the 193 and 171 bands will record a similar absorption \citep[e.g.][]{anzer05}. At the same time, 
the coronal emission  (or intensity)  scale-height in the 193 band is higher than in the 171, implying that at a given height the corona is brighter in the 193 band than in the 171. The coronal emission observed in the  193 band extends to higher altitudes than that observed in the 171 band, and thus it surrounds the prominence more than the 171 coronal emission does. 
Furthermore, we expect that the prominence emission is absent or much lower in the 193 than in  the 171 band. 
As a consequence of these properties we expect the prominence/corona contrast to be higher in 171 than in 193 band, since the latter is the hotter. This is what Figure \ref{fig_aia_prof} shows. The emission observed on top of the prominence in the 171 band, at larger altitude than the coronal scale-height, is due to the PCTR. 

For the 131 band Figure \ref{aia_tresp} shows that the temperatures of the two peaks of sensitivity are different from the typical off-limb coronal temperature, so that the band is intrinsically more blind here. However, we see that the lower temperature profile of the response ( $\log T <$6.5) is quite similar to that of the 171 band but shifted to lower temperature, implying that in the absence of very hot plasma, the corona should look more similar to that observed in the 171 than in the 193. However, it would be weaker and with a shorter coronal scale-height than what is observed in the 171 band. This is confirmed by Figure \ref{fig_aia_prof}.



Our results from simulations, based on the CHIANTI database, show that the 171 passband is always dominated by \ion{Fe}{9}, even in absence of high-temperature TR and coronal plasma. Figure \ref{fig_gts} reinforces this finding. This Figure shows the contribution functions of the main lines in the passband as function of temperature multiplied by the effective area of the filter and the DEM99$\_$55. The same process has been applied to the total contribution function (thick curve), obtained by summing all the lines falling in the passband. The Figure shows that down to $\log T = 5.6$ the intensity of the band, when detected, will still be dominated by the cool wing of the \ion{Fe}{9} contribution function. This happens even though the total contribution function peaks closer to the formation temperature of \ion{O}{6} than to the \ion{Fe}{9}.
 
This finding, together with the information coming from the counts simulated in Table \ref{tab_counts}, suggest that the observed emission in the prominence of Figure \ref{fig_aia} is due to a plasma  temperature of at least about $4\times 10^5~ \mathrm{K}$. This is consistent with the observed   emission in the 131 band above the coronal background. As we said this band is dominated by the emission at similar temperature (\ion{Fe}{8}). The presence of such TR prominence plasma is also consistent with the similarity between the 171 and 304 intensity profiles shown in Figure \ref{fig_aia_prof}. The prominence intensity in the 304 AIA band is dominated by \ion{He}{2} emission \citep{labrosse12} which forms in the temperature range $4<\log T < 5.7$.

Recent theoretical studies lead to similar conclusions. \cite{anzer08}
used a 1D slab model and DEM99 from \cite{parenti07} to derive a temperature profile along the line of sight inside the prominence. They obtained a PCTR reaching values of $4.5 \times 10^5~ \mathrm{K}$ or even higher. This temperature is consistent with the presence of \ion{Fe}{9} emission.
\cite{karpen08} and \cite{luna12} modeled a prominence formation through thermal instability and showed that DEM99 can be closely reproduced up to about $4-5 \times 10^5~ \mathrm{K}$.



However, the finding that \ion{Fe}{9} dominates the 171 band may depend on the  atomic database used. We do not rule out the possibility that non negligible  contributions from cool PCTR lines may be present,  but not detected by our simulations, if such lines are  not included in the CHIANTI atomic database. This may be the case also for the other bands. This database, in fact, is not complete, particularly at low TR and chromospheric temperatures. 
Some of the earlier work on the AIA data  revealed, indeed, some inconsistency between measured and simulated intensities 
 suggesting the presence of unidentified lines in some bands  \citep[e.g.][]{schmelz11}. Recently some progress has been made. For the bands investigated here, \ion{Fe}{7} lines have been identified in the 131 band together with weaker \ion{Fe}{9} 134.08 \AA~ and \ion{Fe}{10} 134.09 \AA~lines \citep[][and references therein]{testa11}. Further investigation is needed to establish their contribution to prominence plasma emission, nevertheless, their presence  would still imply  high-temperature transition region and coronal emission from these structures.

The main contributors to the measured intensities in the AIA bands were identified by previous studies \citep[e.g.][]{odwyer10, martinez11}. In particular, \cite{delzanna11} found that most of 
171 band intensity is due to the emission of \ion{Fe}{9}  when observing the cooler part of warm loops, the  footpoints.
 However, none of these studies investigate  prominences, which are mostly made of chromospheric and low TR plasma. Our results 
 add the information that there is enough PCTR plasma at a temperature able to excite at least \ion{Fe}{8} and \ion{Fe}{9}.

If the nature of prominences PCTR is not very different from the chromosphere-corona transition region, our theoretical results are consistent with TR observations, and unidentified contributions to the AIA intensities may not have much importance. The SOHO/CDS GIS spectrometer  covers the 171 AIA bandpass, and its quiet-Sun spectra  are dominated by the Fe lines, while the oxygen lines do not contribute significantly  (see for example the spectra observed in February 10, 1996 plotted Figure 6 by \cite{harrison97}). This is our finding for a PCTR with plasma at high-temperature transition regions. 


To complete our investigation, we plan to add to this  theoretical investigation an extended analysis of spectroscopic data combined with AIA imaging. In this way it will be possible to observe several of the single contributors to the AIA bands, and infer  the DEM over a wide temperature range. We need, for example, to quantify the opacity level of the lines which, among other things, depends on the mass of the structure. Bigger and more massive prominences are generally observed during the maximum of the solar cycle (implying a stronger magnetic field for their support and stability). This would eventually lead to a difference in the DEM amplitudes and in the prominences opacities over the cycle. The DEMs we used for our simulations were derived from prominences observed in different periods of the solar cycle than  the one observed in 2010 with AIA. It is possible that a solar cycle effect is present.

In conclusion, in the present work we have emphasized the AIA property of being able to detect faint emission in prominences observed with the 171 and 131  bands. Our analysis suggests that this emission comes mainly from \ion{Fe}{9} and \ion{Fe}{8}, implying a PCTR as hot as $4\times 10^5~\mathrm{K}$.




\acknowledgments

SP acknowledges the support from the Belgian Federal Science Policy Office through the ESA-PRODEX programme.
PH was partially supported by the grant P209/12/1652 of the Grant Agency
of the Czech Republic.
LG was supported for this work by a contract from Lockheed Martin to SAO.
CHIANTI is a collaborative project involving George Mason University, the
University of Michigan (USA), and the University of Cambridge (UK).
The AIA data are courtesy provided  by NASA/SDO and the AIA science
team. 








\end{document}